# Similarity Mechanics

Bin-Guang Ma

Email: bgMa@sdut.edu.cn

**Abstract**

The present work provides a new conceptual framework for GUT (Grand Unified Theory) based on a picture of fractal universe. Under a hypothesis of multi-scaled matter structure, we find new clues for the conciliation of quantum and relativity and for the unification of fundamental interactions. A new interpretation for matter wave is proposed as the trajectory of position center of a moving particle with a nucleated structure. The origin of magnetism and gravitation are discussed as the relativistic effects of electrostatic force.

Since 30's of 20th century, human beings' exploring frontier of micro world has been boosted into the field of particle physics. At the beginning of 1960's, thanks to the development of the building technologies of big accelerators, a large quantity of new "elementary particles" has been found. Till then, five forces came into human sight, i.e., electrostatic, magnetic, gravitational, strong nuclear and weak nuclear forces. Among them, electrostatic and magnetic forces had already been unified by Faraday/Maxwell as electro-magnetic force;[1] the electromagnetic force and weak nuclear force were unified through QED, Feynman rules, symmetry, group theory, gauge theory & renormalization *etc.* (by Glashow, Weinberg, Salam, *et al.*) as electro-weak force;[2] on similar lines with quantum-chromodynamics (QCD), the strong force is further considered to be unified, leading to the so-called Standard-Model.[3-5]

Standard Model believes that there exist four fundamental interactions in the nature: strong, weak, electromagnetic and gravitation, and three kinds of elementary particles: gauge-bosons, fermions, and maybe Higgs. Grand Unified Theories (GUT) are just those theories aiming at unifying the different particles and interactions in different existence scales, and String/Membrane theory and its variant Supergravity/quantum-gravity theory are representatives of GUT.[5]

However, there are unsolved problems in these theories which prevent us to reach the ultimate goal of GUT. For example, an unsolved problem of Standard-Model concerns the existence of Higgs particle which is needed for Spontaneous Symmetry Breaking to make particles massive, but till now, there are no such particles experimentally found. As for the string-based theories, they rely on high dimensional space to accomplish the unification of fundamental interactions whilst there is not a consensus of the number of dimension. Another distinct shortcoming of the above theories lies in their long logical path to the end of unification which makes them seem to be too complex to be creditable. The present work provided an alternative logically simple pathway towards GUT based on a picture of fractal universe.

**Basic Principles and Matter Structure**

*Relativity of Measurement* all of our knowledge about quantity are from measurement. Measurement is a procedure of comparison where a measure unit is taken as a standard to be compared with the object to be measured and the quantity of this object is determined according to the number that this object contains the measure unit. The property of measurement that a measure unit is always taken as a reference is called the "relativity of measure". Here measure includes the measure of space, time, speed and energy. The relativity of measure means that big or small (of space), short or long (of time), fast or slow (of speed), high or low (of energy) are all relative. A human's body is smaller than a mountain but bigger than an ant; a human's lifespan is shorter than a tortoise but longer than a fly; a human runs slower than a leopard but faster than a snail; a human has more power than a rabbit but cannot pull back an elephant. All the above examples demonstrate that big or small, long or short, fast or slow and high or low (of energy) are all relative. The key is to see the measure of the object as standard (reference), or namely, the relative scales of the two objects for comparison.

From the "relativity of measure", it can be deduced that the matter structure is infinitely divisible. Proof is as follows: because measure is relative, there is no absolute bigness or smallness; if the matter structure is not infinitely divisible, there must exist absolute bigness or smallness, which contradicts the "relativity of measure"; so the matter structure is infinitely divisible.

### *Basic Principles*

1. Same law principle: all the existence scales keep to the same physical laws;
2. Statistics principle: macro state and micro state coexist.

According to the above "relativity of measure", an infinitely divisible matter structure is obtained. While what is the relationship between matter structures at different existence scales? And what is the relationship between the laws that they keep to? Same law principle tells us that matter structures at different existence scales are similar to each other because they keep to the same physical laws. Then, we can know that if we take human beings' existence scale as the center, there are infinite numbers of "big human" worlds upward and among them the one who takes solar system as an atom is the nearest world from ours along the direction of big scale, and from now on, if we say the "big human world" without otherwise statement, it is default as this world. Likewise, there are infinite numbers of "small human" worlds downward and among them the one who takes an atom as the solar system is the nearest world from ours along the direction of small scale, and from now on, if we say the "small human world" without otherwise statement, it is default as this world. Therefore, the matter structure of our universe is an infinitely divisible fractal.

According to the self-similarity of a fractal, we know that particles at any existence scales have a structure like solar system or atom, i.e., the "nucleated-revolving" structure. At the same time of space scaling, the time is also scaling. That's to say, the time unit used in the small human world is shorter than that of our world whilst the time unit used in the big human world is longer than that of our world. But the scale for space scaling and that for time scaling is not the same, therefore, the light speeds in different scale worlds are not the same. Suppose the light speed in our world is $c$, then the light speed in the small human world is larger than $c$ (may be $c^2$ ) called "fast light" and the light speed in the big human world is less than $c$ (may be $\sqrt{c}$) called "slow light".

However, the human beings in big human world do not feel the so-called (by us) "slow light" in their world is slow because all of the processes in their world are slowed; and thus the speed of "slow light" is still the fastest speed in their world. We say the speed of "slow light" slow just because we take the light speed in our world as a reference. Likewise, the human beings in small human world do not feel the so-called (by us) "fast light" is fast because all the processes in their world are fasted. For an existence scale, the light speed of that scale is the fastest signal speed for that scale which makes the interactions local seen in that scale.

Statistics principle indicates that there are two scales for the description of physical phenomena: macroscopic and microscopic. And there exists a scale transformation able to transform the description of microstate to macrostate. Therefore, the present theory can also be called "Scale Relativity".

**Unification of "Fundamental" Interactions**

*Inverse-Square Law* The form of fundamental interaction is determined by the dimension of the space. The only reasonable form of interaction in 3D space is inverse-square law. Proof is as follows:

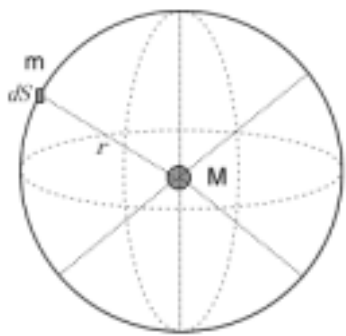

**Figure 1.** Inverse-square force as a result of the decrease of meson flow areal density. The area of the spherical surface $4\pi r^2 \propto r^2$, therefore the areal density of meson flow $\propto 1/r^2$, which is the origin of inverse-square force.

As shown in Figure 1, suppose the interaction between two bodies *M* and *m* are realized by exchanging meson flows; a body's ability of sending and receiving mesons proportions to its matter quantity; and the interaction strength accepted by a body proportions to the meson numbers that it receives.

Then the meson flow sent by body *M* uniformly diffuses to different directions in the 3D space, and then the frontier of this flow is a sphere. Therefore, the area of the front sphere of this meson flow increases with the increase of the square of propagation distance *r*, resulting in the areal density of the meson flow on the front sphere decreases with the square of propagation distance *r*. Then, the meson numbers received by a unit quantity of matter decrease in proportion to the square of *r*. And thus there is:

$$F = K \frac{Mm}{r^2} \qquad (a)$$

This is just the form of inverse-square law where *K* is a constant to be measured by experiment.

An important view of Scale Relativity is that the form of fundamental interaction is determined by the dimension of the space. In the above deduction, three principles are referred: (1) **Locality principle**. That is, the interactions are local ones and realized by exchanging mesons with finite speeds. There is no instant interaction acting at a distance. (2) **Simplicity principle**. That is, the strength of interaction linearly proportions to the matter quantity. (3) **Symmetry principle**. That is, different directions of our space are symmetric with each other and space is isotropic. In 3D space, the formula who keeps to the above three principles is only inverse-square law (a). Similarly, the formula who keeps to the above three principles in 2D space is only the inverse-linear law while that for 4D space must be inverse-cubic law. And so on.

***Gravitation and Electrostatic Force*** They are all inverse-square interactions and essentially the same interaction but appearing at different scales, and thus they are relative. The so-called "gravitation" by us is the electrostatic force for big human; and the so-called electrostatic force by us is gravitation for small human. Therefore, it can be predicted that there must exist anti-gravitation (repulsive force). Galaxies composed of matter and those composed of antimatter must be repulsive to each other so that our solar system is not attracted by other solar systems composed of antimatter and collides and annihilates. For big human, our solar system is just like an atom and sun is the nucleus and the planets are just like electrons. When the planets jump between the orbits running around the sun, slow light is emitted which is the light wave in the big human world and the speed of it is less than $c$ (suppose $c$ is the light speed of our world). Likewise, from the viewpoint of small human, our atom is just like the solar system of their world and the nucleus is the sun and the electrons are the planets. Communication between the nucleus and electrons are like the process of the sun throwing light to the earth. Therefore, the communication between nucleons and electrons is conducted by the "fast light", i.e., the light in small human world.

***Strong and Weak Forces*** Essentially speaking, they are not interactions but reactions. Because strong and weak interactions only manifest in nuclear reaction or particle decay reaction, thus their essence are reactions just like chemical reactions. Because there is only one fundamental interaction— inverse-square interaction —in 3D space which is a long range interaction, the short range properties of strong and weak interactions prove them to be **reactions** not **interactions**. Thinking that strong and weak interactions have equivalent positions to gravitation and electricity to be the fundamental interactions of our world is a wrong thinking. Their short range properties just prove that they are not interactions but reactions just like chemical reactions. Like in chemical reactions where two atoms must approach to each other very close so that the electron orbits of the two reacting atoms are superposed with each other to incur chemical combination, two nucleons can only react with each when they are near enough so that the orbits of the **nucleons' electrons** can superpose with each other (see Figure 2). Just as we do not regard chemical reaction as fundamental interaction of our world, we should not regard nuclear reaction and particle decay reaction as fundamental interaction. By the analogy between particle table and chemical element table, we are convinced that the essence of strong and weak interactions is reaction not interaction.

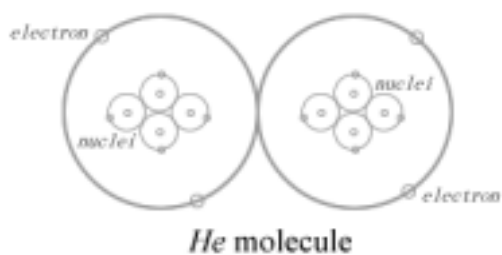

**Figure 2.** Schematic illustration of "nucleus in nucleus" structure of particles.

In a summary of the above, we know that there is only one fundamental interaction in 3D space and that is inverse-square interaction which is a long range interaction. Any short range "interaction" is essentially reaction. In the fundamental interaction, fermions are the agents and bosons are the media. Higgs particles do not exist.

**Conciliation of Relativity and Quantum**

As two cornerstones of the physics building of twenty century, Relativity Theory and Quantum Mechanics have achieved glorious triumph in their own applicable fields, respectively. But there are deep contradictions between them. The contradictions between relativity and quantum mainly reflect at two points: (1) certainty and uncertainty. Relativity is rigorously a theory of certainty. Einstein persists stubbornly in that "God doesn't play dice", while quantum shows some intrinsic uncertainty such as the Heisenberg's uncertainty relation; (2) locality and nonlocality. Relativity is rigorously a local theory and believes that there is the maximum signal speed while quantum demonstrates some nonlocal correlation such as in entanglement.

These diametrically opposed contradictions disturb physicists deeply so that many people believe that one of them must be wrong. However, is this really so? Here we shall tell that the contradictions between relativity and quantum are phenomenological although they seem profound, and relativity and quantum can be unified under the framework of Scale Relativity.

We shall begin with the **uncertainty effect** of quantum. Here we say uncertainty effect instead of uncertainty principle is aiming at emphasizing that the uncertainty is only a kind of effect rather than a principle. Just as space contraction and time dilation in special relativity are effects of **motion relativity**, uncertainty is some kind of effect of **scale relativity**. We cannot determine the position of an electron but we can accurately measure the orbit of a planet. This fact tells us that whether or not we can obtain a precise measure depends on the relative scale of the observer and the object to be observed. From this fact, it can be deduced that small human must be able to accurately measure the position of an electron because an electron looks as big as a planet in their eyes. Likewise, the orbit of our earth must be unable to be accurately measured by big human just like we cannot do it to an electron. Therefore, whether or not an object can be accurately measured is relative, thus, certainty and randomness (uncertainty) is relative. There is no absolute certainty and randomness just as there is no absolute motion and rest. So the uncertainty of quantum does not contradict the certainty of relativity.

Now see the locality and nonlocality. They are also relative. As aforementioned, two particles (such as an electron and a proton) that rest relative to each other communicate by "fast light", *viz.* the light in small human world. Therefore, the interaction between two particles with non-local correlation seen from our world may be local one seen from the small human world. We think the correlation is non-local because the meson speed that connect the two particles surpasses the light speed of our world while small human think the correlation is local because the meson speed that connect them does not surpasses the light speed of their world. Likewise, the local interactions seen from our world may be non-local ones seen by big human. Just because the maximum signal speed at different existence scales are different, the non-local interactions seen from a scale may be local ones when seen from a smaller scale; likewise, the local interactions seen from a scale may be non-local ones when seen from a bigger scale. Therefore, to be local or non-local is relative. There is no absolute locality and nonlocality just as there is no absolute motion and rest. So the nonlocality of quantum does not contradict the locality of relativity and is just some kind of effect of **Scale Relativity**.

Now we see another manifestation of quantum: Discreteness. "Discreteness" and "Continuity" is a pair of relative concepts. That is, a discrete phenomenon seen from a scale may be continuous seen from another scale and vice versa. For example, asphalt road surface is rugged seen by ants but smooth for a van tire. Discreteness is only observable at specific scale. For a bigger scale, it seems continuity due to the rough resolution of the apparatus. For a smaller scale, it is unobservable due to too large span of intervals (even longer than the human activity scope and history of that scale). Therefore, there is no absolute discreteness and continuity just as there is no absolute motion and rest. The difference between discreteness and continuity is only a kind of effect of **Scale Relativity**.

*About Superluminal Speed* For every existence scale, there is a maximum signal speed which is the light speed of that scale. For a scale, the light speed of that scale cannot be surpassed whilst for a smaller scale, there is a faster light speed (the speed of "fast light" seen from the former scale). The smaller the scale is, the faster the light of it runs. Therefore, whether a light speed can be surpassed or not is relative. For every existence scale, it cannot surpass the light speed of its own scale, while for a smaller scale, the light speed of that scale is not a limit. The existence of maximum signal speed is also a kind of effect of **Scale Relativity**.

*About Minimum Quantum of Action* It is believed that Planck constant stipulates the minimum quantum of action of our world. However, because of the **relativity of measure**, for a bigger or smaller existence scale, it is not the minimum quantum of action. The smaller the scale is, the higher the energy density. Therefore, the minimum quantum of action only has a relative meaning; it is only meaningful at a specific existence scale; its existence cannot be taken as a reason for denying the infinitely divisible nature of matter structure.

**The Essence of Matter Wave**

What is matter wave? And what is wave function. Making a clear comprehension for matter wave and wave function in quantum mechanics is another difficult problem disturbing physicists deeply. Here we shall give the answers. The essence of matter wave is the motion trajectory of the position center of a "nucleated-revolving" system in 3D space, i.e., a wave-like motion of position center; and meanwhile it is also the wave-like character of the systematic action of a "nucleated-revolving" system. While the wave function in quantum mechanics is an artificial (man-made) description of the systematic action of the "nucleated-revolving" system as an empirical formula and essentially is an approximation of the wave character of the systematic action in Hilbert space. The module square of wave function reflects the distance between some position in the space (at some time) and the position center of the "nucleated-revolving" system and proportions to the probability to find the

"nucleated-revolving" system at that position. Proofs are as follows:

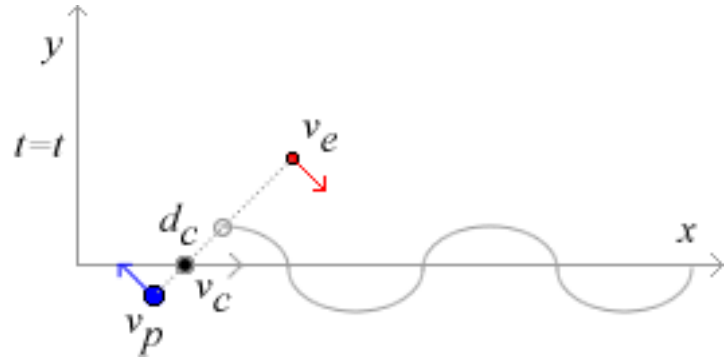

**Figure 3.** The trajectory of the position center of a moving "nucleated-revolving" system is wave.

As illustrated in Figure 3, suppose there is a free "nucleated-revolving" system (a free particle) with a mass center velocity of $v_c$. Here "free" means there is only the interaction between the nucleus and the peripheral particle (internal force) and no external forces acting on the "nucleated-revolving" system. Under such a condition, the system should rotate around their common mass center and the momentum and energy and the angular momentum of this system all conserve in the moving. Suppose the distance between the nucleus and the mass center is $d_p$ and the distance between the peripheral particle and mass center is $d_e$, then the distance between the nucleus and the peripheral particle is $d = d_p + d_e$. Suppose the angular velocity of the system revolving around their mass center is $\omega$, the linear velocity of the nucleus running around the mass center is $v_p$ and the linear velocity of the peripheral particle running around the mass center is $v_e$, then there is $\omega = \dfrac{v_p}{d_p} = \dfrac{v_e}{d_e}$. Taking the direction of the velocity of the mass center as the positive direction, the coordinate frame is established, and then the vertical coordinate $y$ of the mass center is always zero. The position center of the "nucleated-revolving" system is defined as the middle point of the line between the nucleus and the peripheral particle. Suppose the distance between the position center and the mass center of the system is $d_c$, then $d_c = \dfrac{d_e - d_p}{2}$. Suppose the mass of the peripheral particle is $m_e$ and that of the nucleus is $m_p$ and there is $m_p = km_e = km$. Now we deduce the trajectory of the position center of this "nucleated-revolving" system.

Suppose the coordinates of the peripheral particle are $x_e(t)$ and $y_e(t)$. Because the velocity of the mass center $v_c$ is along the positive direction of $x$ coordinate, the trajectory of the peripheral particle is:

$$\begin{cases} x_e(t) = d_e \cos(\omega t) + v_c t \\ y_e(t) = d_e \sin(\omega t) \end{cases} \tag{1}$$

Suppose the coordinates of the nucleus are $x_p(t)$ and $y_p(t)$, then there is:

$$\begin{cases} x_p(t) = -d_p \cos(\omega t) + v_c t \\ y_p(t) = -d_p \sin(\omega t) \end{cases} \tag{2}$$

From (1) and (2), the coordinates of the position center of this system is:

$$\begin{cases} x_c = \dfrac{1}{2}(x_e + x_p) = \dfrac{1}{2}(d_e - d_p)\cos(\omega t) + v_c t \\ y_c = \dfrac{1}{2}(y_e + y_p) = \dfrac{1}{2}(d_e - d_p)\sin(\omega t) \end{cases} \tag{3}$$

Substitute $d_c = \dfrac{d_e - d_p}{2}$ into (3), we get:

$$\begin{cases} x_c(t) = d_c \cos(\omega t) + v_c t \\ y_c(t) = d_c \sin(\omega t) \end{cases} \tag{4}$$

This is the parameter equation of the trajectory of the position center of the "nucleated-revolving" system.

From (4), we know that if the mass of the two parts of the "nucleated-revolving" system (i.e. the nucleus and the peripheral particle) equals, then the position center and the mass center of the system are superposed with each other and $d_c = 0$, and thus there shows no wave character of this system; otherwise, if the mass of the two parts of the "nucleated-revolving" system does not equal to each other, then the position center and the mass center of the system do not superpose with each other and $d_c \neq 0$, and thus the trajectory of the position center is a wave. Generally speaking, the mass of the "nucleated-revolving" system is largely centralized on the nucleus. So the position center does not superpose with the mass center and there shows the wave character of the system.

***Uncertainty Relation*** The distance $d$ between the two parts of the "nucleated-revolving" system is the span of the system in the position space and represents the uncertainty degree of the position of the system; the relative speed $v = \omega d$ is the span of the system in the velocity space and represents the uncertainty degree of the velocity of the system. When the uncertainty degree of the velocity is multiplied by the mass of the system, it represents the uncertainty degree of the momentum of this system. According to the definition and conservation of angular momentum, we get:

$$L = m\omega d^2 = d \cdot m\omega d = \Delta r \cdot \Delta p = \text{constant} . \qquad (5)$$

From (5), we know that the uncertainty of position and the uncertainty of momentum have a relation of one growing and the other declining. That is, the smaller the uncertainty of position, the bigger the uncertainty of momentum; and vice versa. This is just the meaning of Heisenberg's uncertainty relation. Therefore, the uncertainty relation is a result of angular momentum conservation.

Seen from another viewpoint, it is also a result of energy conservation of the "nucleated-revolving" system. Because the inner force of the system is a conservative attractive force, the longer the distance between the two parts of the system (namely, the bigger the uncertainty of position), the higher the percentage of the system's potential energy and the lower the percentage of the system's kinetic energy (namely, the smaller the uncertainty of the momentum); on the contrary, the shorter the distance between the two parts of the system (namely, the smaller the uncertainty of the position), the lower the percentage of the system's potential energy and the higher the system's kinetic energy (namely, the bigger uncertainty of the momentum). In one word, the system's position uncertainty and momentum uncertainty have a relationship that one grows and the other declines.

Readers who are familiar with the deduction procedure of the uncertainty relation from the wave function of quantum mechanics[6] should feel the conciseness of the deduction here. To obtain the uncertainty relation from the wave function of quantum mechanics needs a long and tedious deduction procedure, and the explanations for this relation are even more strange and elusive.[7] Taking the orthodox explanation given by Heisenberg himself as an example, he thinks that uncertainty comes from the disturbance from the instruments to the system to be measured. Later, his this idea has been developed to the function of human's consciousness, sinking into subjective idealism.

From the above deduction, we know that the uncertainty relation is a necessary result of the conservation of the angular momentum (or energy) of a free "nucleated-revolving" system. The deduction procedure is simple and the physical meaning is clear.

***The Wave Character of Action***

By differential operation on equation (4), the velocity of position center is obtained:

$$\begin{cases} v_{g_x} = -d_c \omega \sin(\omega t) \\ v_{g_y} = d_c \omega \cos(\omega t) \end{cases} . \tag{6}$$

Then the action of position center is:

$$\begin{aligned} S_c &= \vec{p} \cdot \vec{r} - E \cdot t = m_c \vec{v}_g \cdot \vec{r}_g - E \cdot t = m_c \left[ v_{g_x} r_{g_x} + v_{g_y} r_{g_y} \right] - \frac{1}{2} m_c v_c^2 t \\ &= m_c \left[ \left( -d_c \omega \sin(\omega t) + v_c \right) \left( d_c \cos(\omega t) + v_c t \right) + d_c^2 \omega \sin(\omega t) \cos(\omega t) - \frac{1}{2} v_c^2 t \right] \\ &= m_c \left[ d_c v_c \cos(\omega t) + d_c v_c \omega (1-t) \sin(\omega t) + \frac{1}{2} v_c^2 t - \frac{1}{2} d_c^2 \omega^2 \right] \end{aligned} \tag{7}$$

where $m_c = (k+1)m$ is the total mass of the whole system.

From (7) it can be seen that the action of the system waves with time $t$. Comparing (7) with Schrödinger's wave function for a free particle:

$$\psi(r,t) = A e^{\frac{i}{\hbar}(p \cdot r - E \cdot t)} = A e^{\frac{i}{\hbar} S}, \tag{8}$$

it can be found that the wave in Schrödinger's equation is an approximation of the wave of the action of a "nucleated-revolving" system in Hilbert space. The original intention of Schrödinger's establishing wave equation is to study the atom structures with a new idea of wave motion. He made an analogy between free particle and plane wave and introduced artificially an imaginary exponential function to make the action of a free particle wave so that the stability of atom structure can be explained by the aid of a concept similar to standing wave. While the cost for this doing is dragging the wave in 3D real space into a mysterious complex space. As for the Planck constant in wave function (8), it is an indication of the precision of our apparatus under the present ability of measurement. Now we have known the essence and origin of matter wave, we do not need Schrödinger's wave equation any longer in principle. Nevertheless, as a set of empirical formulas, the formulism of quantum mechanics is still valuable in dealing with some practical problems.

   ***Distance Function*** Equation (4) is the parameter equation of the trajectory of position center of a "nucleated-revolving" system. For a given time *t*, it gives the average position of the "nucleated-revolving" system in the space. Therefore, we can define a distance function:

$$\begin{aligned} d(x,y,t) &= \sqrt{(x-x_c)^2 + (y-y_c)^2} \\ &= \sqrt{(x(t) - d_c \cos(\omega t) + v_c t)^2 + (y(t) - d_c \sin(\omega t))^2} \\ &\approx \propto 1 - |\psi(x,y,t)|^2 = 1 - \psi^*\psi \end{aligned} \qquad (9)$$

It represents the distance between the position (*x*, *y*) and the position center of the "nucleated-revolving" system and reflects the probability to find the system at position (*x*, *y*) and approximately proportions to the module square of Schrödinger's wave function.

   The position center of a "nucleated-revolving" system represents the position of the whole of this system. Equation (4) shows that there is a certain position of the position center of the system at time *t* and the probability of finding this system at position (*x*, *y*) negatively proportions to the distance from the point (*x*, *y*) to the position center of the "nucleated-revolving" system. Because the nearness or farness of the distance is relative, the probability to find the position center of the "nucleated-revolving" system at some position does not change when the distance function multiplied by a constant (it is equivalent to changing the measure unit of the distance). So the wave function can be normalized.

By now, we know that the relativity of farness or nearness of distance is the foundation that the wave function of quantum mechanics can be normalized and we also know that the distance function is the essence of the target of Born's interpretation for wave function. In addition, we need notice that the position center is just the position center; it only gives the position of the whole of the "nucleated-revolving" system in an average (statistical) sense. In fact, there is neither the nucleus nor the peripheral particle at the position center of the "nucleated-revolving" system. Therefore, by detecting the motion of the whole system via the position center (just as what we do at our existence scale to conduct microscopic experiments), we can only attain a statistical result in the end.

Take hydrogen atom as an example. We can detect the wave character of a hydrogen atom in experiment just because we probe the motion of the position center as the proxy of the whole hydrogen atom. For small humans, they do not probe the position center of a hydrogen atom to describe its motion because in their eyes, a hydrogen atom is as big as a solar system and they can directly determine the position of proton and electron just as we directly probe the position of sun and earth. So what they get is not a statistical result. In a word, that's to say, whether or not obtaining a statistical result is relative. We get the statistical result is some kind of effect of **Scale Relativity**.

***De Broglie Relation, Schrödinger Equation and Born Interpretation*** The concept of matter wave was given by de Broglie in 1924.[8] He was enlightened by the wave-particle duality of photon and guessed that matter particles may also have wave character. His idea was verified by Davisson and Germer in an experiment of electron diffraction in 1927.[9] The introduction of the concept of matter wave by de Broglie is a much-told tale by physicists as a successful example of using analogy. By analogy, de Broglie established the famous relation:

$$\lambda \cdot p = \hbar .  \qquad (10)$$

Comparing (10) with (5), we will find that de Broglie relation is another expression of uncertainty relation actually. In other words, they are the same thing in essence. They are all results of the conservation of angular momentum and energy of the "nucleated-revolving" system.

Schrödinger equation was established by Schrödinger in 1926.[10] It is also a result of analogy. Schrödinger analogized the moving free particle with the propagation of plane wave and established the wave equation named by his name. The concept of wave function originates from his work. While seen in the direct meaning, the wave in Schrödinger's equation and the wave in de Broglie's relation are not the same "wave". For example, the wave length in de Broglie relation has a dimension of length and a direct physical meaning. While the wave length for the wave in Schrödinger equation has no direct meaning because it is a wave in complex space. Therefore, the wave in de Broglie relation and the wave in Schrödinger equation are not the same wave seen from a direct meaning. The essence of the wave in Schrödinger equation is an artificial empirical formula reflecting the wave character of the action of a "nucleated-revolving" system.

Born's interpretation for wave function[11] is called the "Copenhagen" orthodox interpretation. However, the wave in Born's interpretation is a wave of probability (amplitude), and seen from its direct meaning, it is neither the same to the wave in de Broglie relation (whose wave length has a dimension of length) nor the same to the wave in Schrödinger equation (which is the wave of action in complex space). The probability in Born's interpretation essentially reflects the distance from some point of the space to the position center of the "nucleated-revolving" system. Just because of Born's interpretation, the normalization of wave function comes into practice. In fact, no matter for the wave in Schrödinger equation or for the wave in de Broglie relation, there is no need of normalization. Because de Broglie established his relation by analogy with the light wave, just as there is no need of normalization for the light wave, de Broglie's matter wave also does not need normalization. Similarly, Schrödinger established his equation by analogy with ordinary plane wave (such as mechanical wave or electro-magnetic wave), therefore, Schrödinger's wave does not need normalization too just as ordinary plane wave does. However, Born proposes the probability interpretation for wave function in a situation that people have no idea about the origin of the matter wave (i.e., no realization of the ubiquity of the "nucleated-revolving" structure of micro particles) but want to give a unified comprehension of wave-particle duality, thus eliciting the problem of normalization of the wave function.

Summarizing the above, we know that seen from the direct meaning, the wave in de Broglie relation and the wave in Schrödinger equation and the wave in Born's interpretation have different meaning, respectively. However, they are also reflections from different aspects for the same underlying wave: the wave-like trajectory of the position center of the "nucleated-revolving" structure of micro particles. De Broglie relation reflects the conservation of the angular-momentum (or energy) of the "nucleated-revolving" system; Schrödinger equation reflects the wave character of the action of the "nucleated-revolving" system; while the Born interpretation reflects the distance from some point of the space to the position center of the "nucleated-revolving" system.

***Wave-Particle Duality*** From the above deduction, we know that the motion of a "nucleated-revolving" system spontaneously demonstrates the wave character (of the position center) as a whole (the essence of wave is just the propagation of periodicity). On the other hand, the object who can spontaneously demonstrates the wave character must be a "nucleated-revolving" system. Electrons show some stationary-wave-like character on the orbits around the nucleus of an atom; therefore they must have "nucleated-revolving" structures. Generally speaking, any micro particles (except bosons, see explanation later) are "nucleated-revolving" systems.

Now make a comparison with solar system. Earth is running around the sun and the moon is running around the earth. The mass ratio between earth and moon is 80:1, and then the position center of earth-moon system and the mass center of this system do not superpose with each other, and then the trajectory of the position center of the earth-moon system must be a standing wave surrounding at the orbit of the mass center of this system (Figure 4). In addition, our solar system is composed of about ten planets, from which it can be speculated that the sun may be also not a unitary solid ball and that it may be a composite ball made up of about teen ~ twenty small balls as parts of it seeming like an atom nucleus. If really so, some characters of the composite ball may demonstrate in the activities of the sun, for example, the sunspots may be the seams between these part balls, and the precession of perihelion of the planets may originates from the non-uniformity of the gravitation due to the non-uniformity of the density of sun as a composite ball.

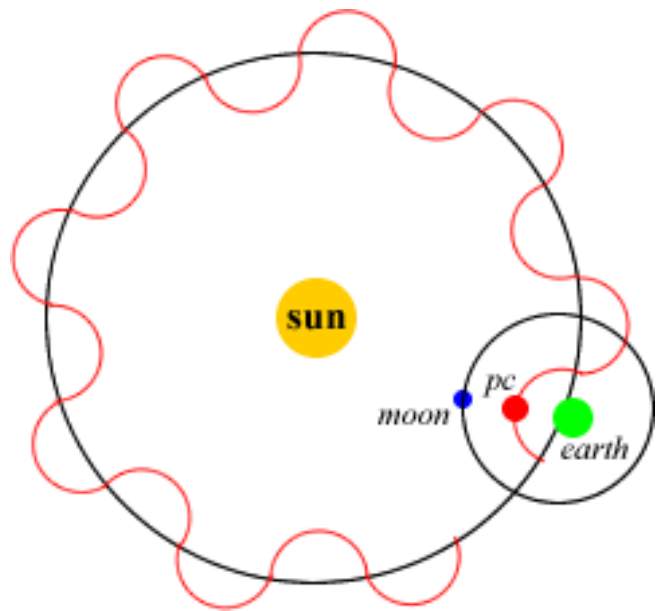

**Figure 4.** Illustration for the trajectory of the position center of earth-moon system.

The essence of matter wave is the wave-like motion of the position center of a "nucleated-revolving" system which embodies the wave-particle duality very well: On one hand, position center demonstrates spontaneously the wave character in the moving; on the other hand, at a specific time, the position center has a specific position so that it also demonstrates the particle character as a whole. However, as aforementioned, position center is just position center; it only represents the position of the whole system in an average (statistical) sense. Therefore, the whole particle only has certain position in a statistical meaning and the results of our experiments remains statistical.

After knowing the essence of matter wave, we can then evaluate different sorts of interpretations for wave function. The following is the evaluation of two representative standpoints. One is Schrödinger's wave package interpretation which emphasizes the wave character too much to face the problem of diffusion of wave package. In fact, from the wave-like motion of the position center, we know that the "nucleated-revolving" system does not diffuse because the attractive inner force between the two parts (nucleus and the peripheral particle). The other interpretation is the ensemble interpretation from Einstein which thinks that the wave character is a collective property of a large quantity of particles which originates from their aggregation. This interpretation emphasizes the particle character too much. In fact, as what we know now, the essence of matter wave is a kind of composite motion (the compound of revolution and translation), and an intrinsic property of "nucleated-revolving" system which can be demonstrated by a single particle and does not rely on the aggregation of many particles. Therefore, the above two interpretations for wave function are all biased in meaning and lose the key point of the truth.

Just as what is said above, because of the **relativity of measure**, whether the wave character is significant or not is also relative. For small human, electrons seem as big as planets, and thus they do not care the wave character of electrons. Likewise, for big human, our earth is as small as an electron, and our solar system is not more than an atom. Therefore, if big humans make a big grating and take a beam of solar systems to throw on it, they will also detect the diffraction phenomenon of solar systems. Hereby, we know that the wave character and particle character are relative, and they are only some kind of effect of **Scale Relativity**.

*Scale Transformation, Symmetry Breaking, Randomness and the Time Arrow*
With the above discussion about matter wave, we can now discuss "Scale Transformation" in more details. "Scale transformation" is actually a simple averaging (summarizing) procedure. Because summarizing operation is a many-to-one map, it is an irreversible procedure which is the origin of symmetry breaking and randomness. The reason is simple: we know 1 + 1 = 2, but if it is asked that 2 equals what? The answer is not necessarily to be 1 + 1, because 1.5 + 0.5 or 0.8+1.2 also equals to 2. That is to say, if knowing the two addends, we can uniquely determine the sum while if knowing the sum, we cannot uniquely determine the two addends. The essence of matter wave may serve as an excellent example to explain that the randomness just originates from "Scale Transformation".

According to "Statistical principle" (special coexistence principle), we know that there are two scales for the description of the state of matter motion: microscopic and macroscopic. At microscopic scale, we can directly describe the motion of the two parts of the "nucleated-revolving" system and need not describe the whole of the system in a manner of position center approximation, therefore uncertainty does not appear. While if because of the limitation of scale, we have to describe the whole of the "nucleated-revolving" system in a manner of position center approximation, then we have to perform a statistical averaging to get the position center of the whole system (usually, this is automatically done by our apparatus which is used for measurement and so we need not do it explicitly) which is an irreversible single direction physical procedure because knowing the positions of the two parts of the "nucleated-revolving" system we can uniquely determine its position center while knowing the position center of the system, we cannot uniquely determine the positions of the two parts, which is the origin of randomness.

The concept of **Time Arrow** is extensively discussed in recently years such as by Stephen Hawking in his "A Brief History of Time"[3] or by Ilya Prigogine in his "The End of Certainty"[12]. The time arrow is just a simple fact: every one of us (except some patients suffering mind diseases) can feel the single direction of time as it is always going from yesterday to today to tomorrow. But why the time is single directional? This problem is not a simple one. To know the answer of this question, we have to refer to the reversibility of physical process. Yet the reversibility of a physical process relies on the determinacy of the physical law that dominates the process. If the physical law is of certainty, just like Newton laws, then the physical process is reversible in principle. Owing to the great success of Newton's mechanics, there prevailed a kind of mechanical world view in eighteen century which believes that the universe is some kind of huge clock. It was said that Laplace had thought that if the position and momentum of each particle are given at a specific time, he can (at least in principle) work out the past and the future of the universe and then time is meaningless in his eyes. Later, due to the appearance and the developments of thermodynamics, it seems to have found the explanation for the time arrow: that is the second law of thermodynamics, i.e. the principle of entropy increase (in an isolated system). If we take the universe as an isolated system (it sounds reasonable because the universe is defined as the totality of our world; then according to this definition, there is nothing outside the universe and if there is, it should be included in the definition), then the single direction of entropy increase may be the underlying reason of the single direction of time. However, this is only a simple correspondence (map); it gives neither the mechanism of time arrow nor the reason why entropy increases.

Here, we shall tell that the increase of entropy and the arrow of time just come from "Scale Transformation". It is generally acknowledged that entropy is a measure of the microstate number of a system and the application of entropy increase principle relies on the adequate randomness of the system, i.e., ergodicity must be satisfied. As aforementioned, randomness just originates from the "Scale Transformation" and this is the precondition of the application of entropy increase principle and the entropy increase principle is only useful for a description of microstates at a macroscopic level. If we can directly measure and control the system at microscopic level (as if we have the ability of Maxwell's demon), then entropy is a useless concept for us (Note: if we do not care a more microscopic existence scale) and every process is reversible for us and thus there is no single directional time in our eyes.

Summarizing the above, we know that there is no absolute reversibility or irreversibility for a physical process and that there is no absolute certainty or randomness of physical phenomenon. Whether a physical process reversible or not and whether a physical phenomenon certain or random rely on the measure scale. We detect the motion of an electron to be random but for small human, the motion of an electron is completely certain no matter for position or momentum. So whether or not time is reversible are relative and the time arrow is a kind of effect of Scale Relativity.

**The Essence of Field**

*What is Field?* Field is meson flow. And the force lines are intuitive description of field.

*What is Magnetic Field?* Magnetic field is a kind of measure effect of "Motion Relativity" (Einstein's special relativity) coming from the space contraction.

As shown in Figure 5, suppose the interactions between the protons in $A_1$ and the protons in $A_2$ is $F_{p1}^{p2}$ and the interactions between the electrons of $A_1$ and those of $A_2$ is $F_{e1}^{e2}$ and the interactions between the protons in $A_1$ and the electrons in $A_2$ is $F_{p1}^{e2}$ and the interaction between the electrons in $A_1$ and the protons in $A_2$ is $F_{e1}^{p2}$. Then,

$$F_{attractive} = F_{p1}^{e2} + F_{e1}^{p2}, \qquad F_{repulsive} = F_{p1}^{p2} + F_{e1}^{e2} \qquad (11)$$

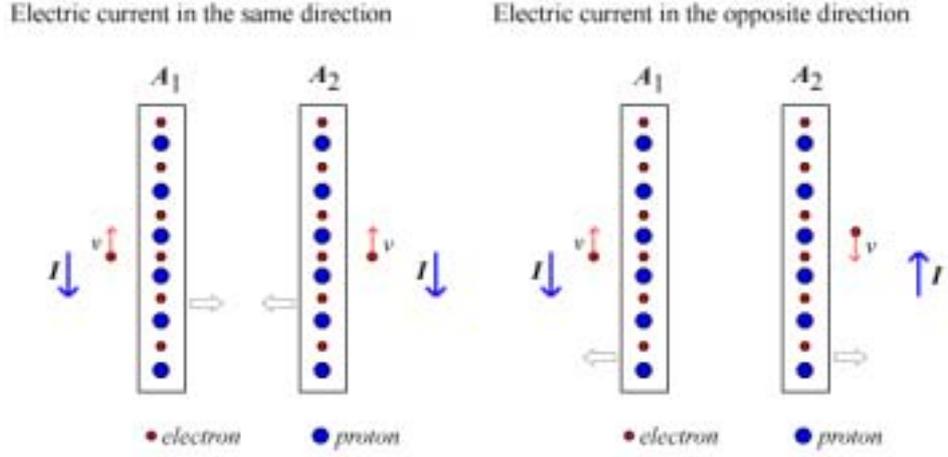

**Figure 5.** Schematic illustration for the generation of magnetic filed. When the electric currents in the two conductors $A_1$ and $A_2$ are in the same direction, the two conductors attract each other and when the electric currents in them are in the opposite direction, they repulse each other.

Before adding voltage, because there are equal positive and negative charges in the two conductors, there is

$$F_{p1}^{p2} = F_{e1}^{e2} = F_{p1}^{e2} = F_{e1}^{p2}, \qquad (12)$$

and thus

$$F_{attractive} = F_{repulsive}. \qquad (13)$$

After charging, the electrons in the two conductors begin directional movements (without losing generality, suppose the magnitudes of the two currents in the two conductors are equal to each other).

Firstly, we see the situation where the two currents in the two conductors are in the same direction. In this case, the protons in the two conductors are resting relative to each other and the electrons in the two conductors are resting relative to each other too. Therefore, $F_{p1}^{p2}$ and $F_{e1}^{e2}$ do not change, and thus the repulsive force between the conductors $F_{repulsive} = F_{p1}^{p2} + F_{e1}^{e2}$ also does not change, i.e.,

$$F'_{repulsive} = F_{repulsive} = F_{p1}^{p2} + F_{e1}^{e2} \qquad (14)$$

where $F'_{repulsive}$ is the repulsive force after charging.

The electrons in $A_2$ have a directional speed $v$ relative to the protons in $A_1$, and then because of the conservation of the quantity of electricity (i.e., the quantity of electricity is an invariant of "motion relativity") and the contraction effect of length (of the "motion relativity"), the areal number density of meson flow emitted from the protons in $A_1$ and received by the electrons in conductor $A_2$ seems increased and as a consequence $F_{p1}^{e2}$ increased. The following is a concise deduction procedure.

According to length contraction formula of "motion relativity",

$$L = L_0\sqrt{1-\frac{v^2}{c^2}} \cong L_0\left(1-\frac{1}{2}\frac{v^2}{c^2}\right) = L_0\left(1-\frac{1}{2}\beta^2\right) \qquad (15)$$

there is

$$\Delta L = L - L_0 = -\frac{1}{2}\beta^2 L_0. \qquad (16)$$

Let

$$L\downarrow = \left|\frac{\Delta L}{L_0}\right| = \left|-\frac{1}{2}\beta^2\right| = \frac{1}{2}\beta^2. \qquad (17)$$

Then "$L\downarrow$" is the decrease factor of length.

According to the definition of the areal number density of meson flow, there is

$$\because D_a = \frac{Q}{A} = \frac{Q}{LL_0} \text{ and } L\downarrow = \frac{1}{2}\beta^2$$
$$\therefore D_a\uparrow = \frac{1}{2}\beta^2 \qquad (18)$$

where "$D_a\uparrow$" is the increase factor of the areal number density of meson flow.

Therefore,

$$F_{p1}^{e2}\uparrow = \frac{1}{2}\beta^2 \text{ and } F_{e1}^{p2}\uparrow = \frac{1}{2}\beta^2 \qquad (19)$$

and then

$$F_{attractive} \uparrow = \left(F_{p1}^{e2} + F_{e1}^{p2}\right) \uparrow = \frac{1}{2}\beta^2. \tag{20}$$

As a result,

$$F'_{attractive} = F_{attractive}(1 + \frac{1}{2}\beta^2) = F_{attractive} + \frac{1}{2}\beta^2 F_{attractive} > F'_{repulsive} \tag{21}$$

where the $F'_{attractive}$ is the attractive force after charging. Then we detect the phenomenon of a net attraction between the two conductors which looks like there is magnetic field between the two conductors.

Secondly, we see the situation where the two currents in the two conductors are in the opposite direction. In this case, the protons in the two conductors are resting relative to each other and then, $F_{p1}^{p2}$ does not change. The relative speeds of electrons in the two conductors are $2v$ (Here Galileo velocity addition law is used because the speeds of the directional motion of electrons in conductors are not very high), then the length contraction factor is

$$L \downarrow = \left|\frac{\Delta L}{L_0}\right| = \left|-\frac{1}{2}\left(\frac{2v}{c}\right)^2\right| = 2\beta^2 \tag{22}$$

Then, there is $F_{e1}^{e2} \uparrow = 2\beta^2$.

Therefore,

$$F'_{repulsive} = F_{p1}^{p2} + F_{e1}^{e2} = F_{p1}^{p2} + \left(1 + 2\beta^2\right)F_{e1}^{e2} = F_{repulsive} + 2\beta^2 F_{e1}^{e2} = F_{repulsive} + \beta^2 F_{repulsive}.$$

Consider

$$F'_{attractive} = F_{p1}^{e2}\left(1 + \frac{1}{2}\beta^2\right) + F_{e1}^{p2}\left(1 + \frac{1}{2}\beta^2\right) = F_{attractive} + \frac{1}{2}\beta^2 F_{attractive}$$

and

$$F_{attractive} = F_{repulsive},$$

we arrive at

$$F'_{repulsive} > F'_{attractive} \qquad (23)$$

Then we shall detect a net repulsion between the two conductors which are explained as the magnetic field between the two conductors.

From the above, we know that the magnetic field is a kind of effect of "motion relativity" (special relativity) of electric field, namely, the motion of electricity generates the magnetism. There is never magnetism without electricity. So, magnetic charge and magnetic monopole do not exist.

***What is Gravitation?*** Gravitation is the most common force around us. According to Einstein's General Relativity, it can be regarded as the curvature of space-time[13, 14]. In 2005, Prof. R. C. Gupta (I.E.T., Lucknow, India) proposed an alternative explanation that gravity can be regarded as the second-order relativistic manifestation of electrostatic force[15]. The following figure and table are recreated according to Prof. Gupta's ideas.

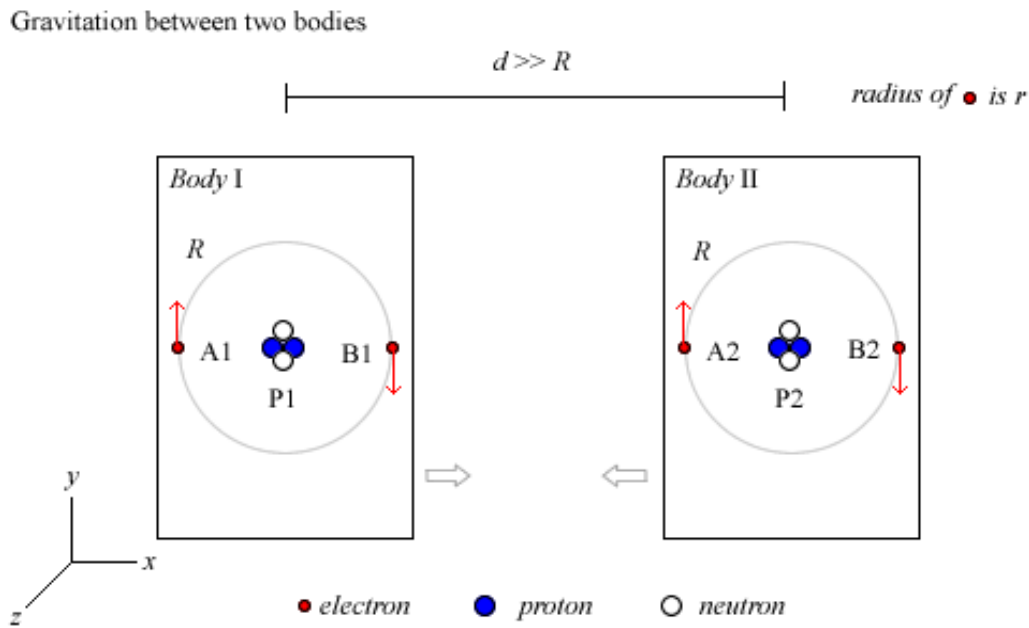

**Figure 6.** Schematic illustrations for the generation of gravitational field. Suppose two bodies (body I and body II) contain Helium-like atoms; the distance *d* between the two bodies is far larger than the radius of the atoms *R*. For simplicity, suppose two atoms (each one in each body) are arranged as shown where A1 & B1 are electrons in Body I and A2 & B2 are electrons in body II; the two protons in body I are grouped as P1 and the two protons in body II are grouped as P2. The net Attraction (+) and Repulsion (-) factors between electrons and protons of the two atoms in the two bodies due to length contraction are listed in the following Table.

**Table 1** Net Attraction (+) and Repulsion (-) factors between electrons and protons of the two atoms in the two bodies due to length contraction

| Observations | Net Attraction (+) or repulsion factor due to length contraction | |
|---|---|---|
| | **Without relativistic velocity addition (first order effect)** | **With relativistic Velocity addition (second order effect)** |
| A1 as observer sees A2 | 0 $\quad = 0$ | 0 $\quad = 0$ |
| A1 as observer sees P2 | $+2 \times (1/2) v^2/c^2 = +\beta^2$ | $+2 \times (1/2) v^2/c^2 = +\beta^2$ |
| A1 as observer sees B2 | $(-1/2) \times (2v)^2/c^2 = -2\beta^2$ | $(-1/2) \times (2v)^2/c^2 \{1/(1+v^2/c^2)^2\} = -2\beta^2(1-2\beta^2)$ |
| B1 as observer sees A2 | $(-1/2) \times (2v)^2/c^2 = -2\beta^2$ | $(-1/2) \times (2v)^2/c^2 \{1/(1+v^2/c^2)^2\} = -2\beta^2(1-2\beta^2)$ |
| B1 as observer sees P2 | $+2 \times (1/2) v^2/c^2 = +\beta^2$ | $+2 \times (1/2) v^2/c^2 \quad = +\beta^2$ |
| B1 as observer sees B2 | 0 $\quad = 0$ | 0 $\quad = 0$ |
| P1 as observer sees A2 | $+2 \times (1/2) v^2/c^2 = +\beta^2$ | $+2 \times (1/2) v^2/c^2 \quad = +\beta^2$ |
| P1 as observer sees P2 | 0 $\quad = 0$ | 0 $\quad = 0$ |
| P1 as observer sees B2 | $+2 \times (1/2) v^2/c^2 = +\beta^2$ | $+2 \times (1/2) v^2/c^2 \quad = +\beta^2$ |
| **atom-I as observer sees atom-II** | Total $\quad = 0$ | Total $\quad \approx + 8\beta^4$ |

As shown in Figure 6 and Table 1, the net attraction will appear when the Lorentz velocity addition law is used in the length contraction formula and Newton's gravitational formula can be derived from Coulomb's electrostatic force formula to be:

$$F = \{1/(4\pi\varepsilon)\} \cdot q_1 [q_2 f] / d^2$$
$$= \{1/(4\pi\varepsilon)\} \left(\tfrac{1}{2} N_A m_1 e\right) \left[\tfrac{1}{2} N_A m_2 e \cdot \{K(v/c)^{n_1} (r/R)^{n_2}\}\right] / d^2 \qquad (24)$$
$$= G \cdot m_1 m_2 / d^2$$

and the gravitational-constant $G$ can be theoretically estimated as:

$$G = \{(\tfrac{1}{2} N_A e)^2 / (4\pi\varepsilon)\} \cdot \{K(v/c)^{n_1} (r/R)^{n_2}\}. \qquad (25)$$

The detailed procedure for the deduction and estimation of these formulas are presented in Gupta's paper[15].

According to Prof. Gupta, the "Van der Waals" force between two mesoscopic bodies such as two molecules, which are due to charge distribution, can be regarded as the "zeroth-order" relativistic effect of electrostatic force, meaning with "no" relativistic effect at all; the magnetism can be regarded as "first-order" relativistic manifestation of electrostatic force because it considers the length contraction of special relativity; and the interactions between two macroscopic bodies can be regarded as the "second-order" relativistic manifestation of electrostatic force of special relativity because it considers the relativistic velocity addition (Lorentz velocity addition law). This view can conceptually unify the (long range) interactions as electrostatic force and its relativistic effects. But there are residual problem of this view because the gravitation between two neutrons cannot be explained in the way of two atoms unless that neutrons can also be regarded as a nucleated structure like atoms. According to Scale Relativity, we know that all the (real) particles have a "nucleated-revolving" structure. Then the forces can be unified as the inverse-square interaction and its relativistic effects, which fulfills the unification of long range interactions.

***Does the mass really increase?*** The mass of a body will increase with the increase of the body's speed is an important inference of special relativity (motion relativity). But is this the truth? Up to now, our evidence about the increase of mass mainly comes from the acceleration experiments where it is found that it becomes more and more difficult to accelerate particles and when it is calculated using the formula of charge-mass ratio and seems as if the mass has increased. However, what about the truth? This should be started from the mechanism of acceleration. When a particle is accelerated in the external field, the increase of its momentum comes from the impulsive force of meson flow. Figuratively, it seems like a cobble is pushed rolling by the flow of a brook. The rolling speed of the cobble cannot surpass the speed of the stream. And when the speed of the cobble becomes faster and faster, the difference between the speeds of the cobble and the stream becomes smaller and smaller, and then the impulsive force the cobble received becomes less and less, and as a result, it seems that the cobble becomes more and more difficult to be accelerated. If we take the impulsive force as an invariant, we will think that the mass of the cobble is increased. While in fact, the mass of the cobble does not change, and just the impulsive force becomes less and less. Similar situation appears for particles to be accelerated. Therefore, we know that mass just like electricity is also an invariant of "motion relativity". Previously, thinking the increase of the mass with the speed is a misunderstanding.

*Electricity and Mass* Essentially, mass and electricity are the same physical quantity: inverse-square quantity. They are the same physical quantity manifested at different existence scales. If the matter is not infinitely divisible, then if electricity is discrete, mass is discrete. In fact, as aforementioned, the matter is infinitely divisible, so discreteness and continuity is relative, just a kind of effect of **Scale Relativity**. The discreteness of charge originates from our observation scale and the inadequate resolution of our apparatus. In fact, fractional charge has been suggested in QCD. This has already challenged the traditional view of integral discreteness of electricity.

It can be reasonably expected that the relativity of discreteness and continuity will be recognized with the development of the resolution of our apparatus.

*What is Spin?* Spin is the self-rotation of a particle and at the same time, is the revolution of the peripheral particle on its orbit around the nucleus. Therefore, the angular momentum of spin and the angular momentum of orbit of peripheral particle are relative and up to our observation scale. What is spin (the angular momentum of self-rotation) seen from the macroscopic scale where the nucleated structure of the particle is out of sight may be the angular momentum of orbit of the peripheral particle when seen from microscopic scale where the nucleated structure of the particle is in sight; and vice versa. This is why they keep to the same commutation relation. Spin, as a kind of angular momentum, reflects the polarity of a particle or more accurately speaking, the directional property of a particle. Why we always get two values of spin (a positive one and a negative one) when we detect the spin of fermions? The reason is as follows. Although, as angular momentum, the spin of a particle can have many directions, but when coupling with each other as magnetic moment, there are only two arrangement states of equilibrium: up-magnetic (paramagnetic) and down-magnetic (diamagnetic). The former is a stable equilibrium and has lower energy; and the latter is an unstable equilibrium and has higher energy. That is why we can only detect two values of spin.

*What is isospin?* It is a wrong concept introduced in a wrong way. It has its historical meaning in a phenomenological theory staying in an empirical stage, but it cannot be taken as a concept with the meaning of truth. Once we have recognized the true structure of matter, we should abandon it in time.

***Super-Symmetry*** Super-symmetry is a great concept. It is proposed originally for the unification of the so-called fundamental interactions. It aims to construct a bigger group to accommodate both bosons and fermions, i.e., letting them be the representation of this group. Regarding particles as representations of groups is a beautiful idea and a model of symmetry guiding physical research. But symmetry is a sward with two sharp edges: It can guide physical research in the right way and also can misguide it. Which result appears depends on whether it is used correctly. If someone wants to find a group to accommodate all the chemical molecules, i.e., rendering them to be the representations of this group, his doing is reasonable or unreasonable just as we render the so-called elementary particles to be the representations of some group. Anyhow, the doing of correlating the group representation with the law of fundamental interaction is unreasonable. The law of fundamental interaction is determined by the dimension of space and has nothing to do with group representation.

***String and Roton*** Representing particles as the different vibration model of strings is a beautiful idea in string theories[16] and can be appreciated as a piece of artwork but cannot be regarded as the truth because there is no solid foundation of this idea. In contrast, "nucleated-revolving" system (called "Roton" for short) is a model with solid foundation that at least the structures of atom and solar system are all this kind of structures. The idea that all the real particles (namely fermions) are Rotons satisfies the invariance of scale transformation and can unify interactions. Logically speaking, there are no more than two forms of interaction: "direct contact" or "via media". In 3D space, the sole reasonable interaction in the form of "via media" (long range interaction) is the inverse-square interaction whose formula has been proved above to be determined by the dimension of the space, and Rotons are the nucleated revolving structures naturally formed under such an interaction. While all the interactions in the form of "direct contact" can be regarded as the combination and decomposition of Rotons.

***Roton and Mass Point*** Roton, as a concept, represents the ubiquitous "nucleated-revolving" structure of particles. It differs from Newton's mass point model in that the mass point model is only a geometric point without inner structures and only with a man-set property: mass, while Roton has a recursive, infinitely divisible "nucleated-revolving" structure. Mass point can be regarded as a far distance approximation of Roton because when observed from a far distance, plenty of inner structures of Rotons are unobservable to the observer due to the distance. But when observed from a near distance, the details of the Roton structure are not neglectable. The Roton model has some advantages compared with mass point model because it avoids many singularities resulting from the zero volume of a geometric point representation but still can be taken as a point when the cared scale are far larger than the scale of the diameter of the Roton.

**Acknowledgement**

Many Thanks to Hong-Yu Zhang, Wu-Sheng Dai and Gang Lv for valued discussions.